\documentclass[twocolumn,times,twocolappendix]{aastex631}

\usepackage{amssymb,amsmath}
\usepackage{natbib}
\usepackage{longtable}
\shorttitle{Intracluster magnetic fields}
\shortauthors{J. Xu and J. L. Han}

\begin{document}

\title{Evidence for strong intracluster magnetic fields in the early Universe}

\email{xujun@nao.cas.cn, hjl@nao.cas.cn}

\author{J. Xu}
\affil{National Astronomical Observatories, Chinese Academy of Sciences,
     A20 Datun Road, Chaoyang District, Beijing 100101, China}
\affil{CAS Key Laboratory of FAST, NAOC, Chinese Academy of Sciences}

\author{J. L. Han}
\affil{National Astronomical Observatories, Chinese Academy of Sciences,
     A20 Datun Road, Chaoyang District, Beijing 100101, China}
\affil{CAS Key Laboratory of FAST, NAOC, Chinese Academy of Sciences}
\affil{School of Astronomy and Space Sciences,
       University of Chinese Academy of Sciences, 
       Beijing 100049, China}

\begin{abstract}
  The origin of magnetic fields in clusters of galaxies is still a
  matter of debate. Observations for intracluster magnetic fields over
  a wide range of redshifts are crucial to constrain possible scenarios
  for the origin and evolution of the fields. { Differences of
  Faraday rotation measures (RMs) of an embedded double radio sources,
  i.e. a pair of lobes of mostly Fanaroff--Riley type II radio
  galaxies, are free from the Faraday rotation contributions from the
  interstellar medium inside the Milky Way and the intergalactic
  medium between radio galaxies and us, and hence provide a novel way
  to estimate average magnetic field within galaxy clusters.} We have
  obtained a sample of 627 pairs whose RMs and redshifts are available
  in the most updated RM catalogues and redshift databases. The RM
  differences of the pairs are derived. { The statistically large
    RM differences for pairs of redshifts $z>0.9$ indicate that
    intracluster magnetic fields is as strong as about 4~$\mu$G. Such
    strong magnetic fields in the intracluster medium at the half age
    of the Universe, comparable to intracluster field strength in
    nearby galaxy clusters, pose a challenge on the theories for
    origin of cosmic magnetic fields.}
\end{abstract}


\section{Introduction}
\label{sect1}

In the past decades, magnetic fields in galaxy clusters have been
observed and studied \citep[see review
  of][]{han17,ct02,gf04,fgs+08,fgg+12}. The magnetic fields are
crucial for a comprehensive understanding of radio emission from the
diffuse intracluster medium (ICM). The presence of diffuse radio halos
and radio relics in galaxy clusters is the direct evidence for
magnetic fields in the ICM \citep[e.g.][]{gbf+09,vrbh10}. Under the
minimum energy hypothesis or equipartition approach, magnetic fields
permeating the ICM are roughly estimated from the radio emission
intensity maps with a strength of a few micro-Gauss
\citep[e.g.][]{gf04}.

Statistical study of Faraday rotation measures (RMs) of radio sources
within or behind galaxy clusters is an alternative way to investigate
magnetic fields in galaxy clusters
\citep[e.g.][]{ktk91,ckb01,gdm+10,bfm+10,bvb+13,pjds13,bck16}. When a
linearly polarized electromagnetic wave signal travels through a
magnetized plasma, the plane of polarization is rotated by an angle
$\Delta \psi$ proportional to the wavelength squared $\lambda^2$, i.e.
\begin{equation}
\Delta \psi = \psi-\psi_0= \rm{RM} \cdot \lambda^2,
\end{equation}
where $\psi$ and $\psi_0$ are the measured and intrinsic polarization
angle, and RM is the rotation measure which is an integrated quantity
of the product of the thermal electron density $n_e$ and magnetic
field strength ${ B}$ from the source to us, most effectively
probing the fields along the line of sight.  For a polarized radio
source at redshift $z_{\rm s}$, RM is expressed by
\begin{equation}
  {\rm RM} = 812\int_{\rm source}^{\rm us} n_e { B} \cdot d{ l}
  =812 \int_{\rm z_s}^{\rm us}\frac{n_e(z)B_{||}(z)}{(1+z)^{2}}\frac{dl}{dz} dz.
\label{rmz}
\end{equation}
The electron density $n_e$ is in cm$^{-3}$, the magnetic field is a
vector ${ B}$ (and magnetic field along the line of sight $B_{||}$)
in units of $\mu$G, and $d{ l}$ is the unit vector
along the light path towards us in units of kpc. The comoving path
increment per unit redshift $\frac{dl}{dz}$ is in kpc and $(1+z)^2$
reflects the change of wavelength at redshift $z$ over the path transformed
to the observer's frame.

The observed rotation measure ${\rm RM}_{\rm obs}$, is a sum of the foreground
Galactic RM (GRM) from the Milky Way, the rotation measure from
intergalactic medium ${\rm RM}_{\rm IGM}$ and intrinsic to the source
${\rm RM}_{\rm in}$
i.e.
\begin{equation}
  {\rm RM}_{\rm obs} = {\rm GRM} + {\rm RM}_{\rm IGM} + {\rm RM}_{\rm in}.
\label{rmobs}
\end{equation}
When studying RMs of sources at a cosmological distance, one has to
account RM contributions from all kinds of the intervening medium
along the line of sight. For most extragalactic radio sources, the
foreground Galactic RM is the dominant contribution. If the foreground
GRM is not assessed properly, it is impossible to get small
extragalactic contributions. There have been many efforts to
investigate the foreground GRM \citep[e.g.][]{hmbb97, ojr+12, xh14,
  ojg+15}. RM values intrinsic to a radio source (${\rm RM}_{\rm in}$)
at a redshift of $z_{\rm s}$ are reduced by a factor
$(1+z_{\rm s})^{-2}$ due to change of $\lambda$ when the values are transformed
to the observer's frame. The typical distribution of source-intrinsic
RMs of distant quasar-like sources is only several rad~m$^{-2}
$\citep{bsg+14}. The RMs from the intergalactic medium ${\rm RM}_{\rm
  IGM}$ may have several contributors, such as rotation measures from
the cosmic webs, intervening galaxy halos and intracluster medium on
the line of sight.  The rotation measure from the cosmic webs might be
traced by Ly$\alpha$ forest, and there have been some simulations on
their contribution \citep[e.g.][]{bbo99,ar10,ar11,ptu16}. It is very
small ($\sim$1--2 rad~m$^{-2}$) that it hardly be detected from
present available data \citep{xh14b,omv+19,obv+20}. The excess of
rotation measure from galaxy halos or protogalactic environments has
been studied by intervening absorbers like Mg~II absorption lines
\citep[e.g.][]{bml+08,foc+14,frg+17}. \citet{jc13} and \citet{mcs20}
obtained an increase in the distribution deviation of around 8
rad~m$^{-2}$ for quasars with Mg~II absorption lines. Statistics of
RMs of polarized radio sources located inside or behind galaxy
clusters \citep[e.g.][]{ktk91,ckb01,gdm+10,bfm+10,bvb+13,pjds13,bck16}
show the RM excess for the contributions from the intracluster medium
with an amplitude from a few to a few tens of rad~m$^{-2}$
\citep{xh14b,gdm+10,ckb01}.

It is now well established that the magnetic fields are ubiquitous in
the ICM \citep[e.g.][]{ct02}.  The intracluster magnetic fields are
dominated by turbulent fluctuations over a range of scales. The field
strength decreases from the central regions to the outskirts. The 
spatial power spectrum is well represented by a Kolmogorov power
spectrum \citep{bfm+10}. Turbulent magnetic fields with a coherence
length of a few kpc are indicated by RM dispersion studies of
polarized radio sources \citep[e.g.][]{ktk91,gdm+10} and found in both
relaxed clusters and merging clusters regardless of dynamic states
\citep{ckb01,bck16,sd19}.  Coherent rotation measures of radio relics
reveal large-scale ($>$100 kpc) compressed magnetic fields
\citep{ore+14,kbh+17}.  The organized magnetic fields are responsible
for systematic RM gradient over lobes of radio galaxies
\citep[e.g.][]{tp93}.  The ordered net magnetic fields can be
considered as the large-scale fluctuations at the outer scale of
turbulent magnetic fields where the energy is injected
\citep{vmg+10}. The magnetic fields close to the center of galaxy
clusters are more disturbed and tangled with a strength of a few
micro-Gauss while those near the outskirts are more representative for
the large-scale fluctuation component with a field strength of an
order of magnitude smaller \citep{rkcd08}.

\begin{figure}
\centering
\includegraphics[angle=0,width=80mm,trim=50 100 0 50,clip]{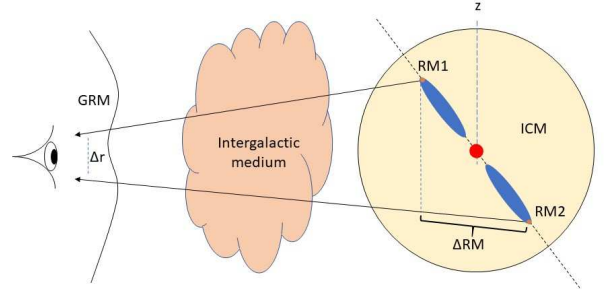} 
\caption{A schematic diagram showing a pair of lobes from a FR~II
  radio galaxy with observed RMs on each side (RM1 and RM2,
  respectively).  The RMs of the pair with such a small angular
  separation ($\Delta$r) of an order of arcminutes have almost the
  same Galactic contributions (in general coherent from several
  degrees at low Galactic latitudes to tens of degrees at high
  Galactic latitudes) and the same intergalactic contributions in
  front of the lobes. Therefore the RM difference ($\Delta$RM) between
  the two lobes are the best probes for the magnetic properties of the
  ICM.}
\label{FRIIsch}
\end{figure}

The RM difference of a pair of lobes from an embedded FR~II radio
galaxy \citep{fr74} are the best probes for the magnetic fields in the
ICM and their redshift evolution, because both the foreground Galactic
RM and the RM contributions on the way to the cluster in all
intervening galactic and intergalactic medium can be diminished, as
depicted in Figure~\ref{FRIIsch}. The real physical pair of lobes are
the bulk of radio emission from a galaxy on opposite sides, formed
when central active galactic nuclei produce two opposite collimated
jets that drive relativistic electrons running in magnetic fields into
the lobes to generate synchrotron emission \citep{br74}. The environs
of the host galaxy must be rich of gas. The jets travel through the
interstellar medium of the host galaxy, and stay supersonic to a great
distance to push their way through the external medium where a shock
front is formed as shown by hot spots. The end of the jets move
outwards much more slowly than material flows along the jets. A back
flow of relativistic plasma deflected at the end of the jets forms the
lobes. The gaseous environment they inhabit is very important to
provide a working surface for the jets to terminate, therefore, the
ICM provides an ideal environment for producing FR~II radio
sources. The observed radio radiation from FR~II type radio sources is
often highly linearly polarized \citep[e.g.][]{bfm+10}. { The
  Laing-Garrington effect strongly suggests the existence of
  intracluster magneto-ionic material surrounding the radio sources
  causing asymmetry in the polarization properties of double radio
  sources with one jet \citep{lai88,glcl88}.} Many double radio
sources have been detected from galaxies at low redshifts ($z<0.3$),
and a large number of sources have been found in dense cluster-like
gaseous environments at higher redshifts
\citep{ymp89,hl91,wd96,pvc+00,md08}.

It is not known if there is any evolution of intracluster magnetic
fields at different cosmological epochs. Statistical studies of the
redshift evolution of {\it net} rotation measures contributed by the
ICM is the key for the puzzle.  { Cosmological simulations by
  \citet{ar11} predicted the redshift dependence of extragalactic
  rotation measures caused by the intergalactic medium. Contributions
  by galaxy clusters, however, could not be properly modeled given the
  cell size in their simulations.}  Previously, there have been a
number of works to investigate the redshift evolution of extragalactic
rotation measures \citep{hrg12,nsb13,xh14b,ptu15,lrf+16,opa+17}, which
were generally made for the whole contributions on the path from the
observer to the sources. A marginal dependence of redshift was
found. In the early days, the RM differences were also studied for a
small number of double radio galaxies at low Galactic latitudes to
investigate the enhanced turbulence in the interstellar medium
\citep{sc86,prm+89,lsc90,ccsk92,ms96}.  \citet{akm+98} studied 15
radio galaxies at high redshift $z>2$ with large rotation measures,
and claimed their RM contributions are likely to be in the vicinity of
the radio sources themselves. \citet{gkb+04} and \citet{opa+17}
concluded that no statistically significant trend was found for the RM
difference of two lobes against redshift.  \citet{vgra19} classified a
large sample of close pairs and found a significant difference of
$\sim$5--10 rad~m$^{-2}$ between physical pairs (separate components
of a multi-component radio galaxy or multiple RMs within one of the
components) and random pairs, though the redshift dependence of the
physical pairs is not evident. \citet{obv+20} used a similar method
but high precision RM data from the LOFAR Two-Metre Sky Survey, and
they find no significant difference between the $\Delta$RM
distributions of the physical and non-physical pairs. In fact, the
uncertainty of RM measurement is a very important factor for the
evolution investigation. For example, very small RM differences
(1$\sim$2 rad~m$^{-2}$) between the lobes of large radio galaxies at
low redshifts can be ascertained with high precision observations
\citep{omv+19,bowe19,sob+20}. RM differences for a larger sample of
pure double radio sources is necessary to further investigate their
correlation with redshift.

A real pair of two physically associated lobes shown as double radio
sources have a small separation and an almost the same flux density,
which can be found in the Jansky Very Large Array (JVLA) Sky Survey
\citep[NVSS;][]{ccg+98}.  \citet{tss09} have reprocessed the 2-band
polarization data of the NVSS, and obtained the two-band RMs for
37,543 sources. \citet{xh14} compiled a catalog of reliable RMs for
4553 extragalactic point radio sources. In addition to the previously
cataloged RMs, many new RM data are published in the literature. In
this paper, we have classified RM pairs in the NVSS RM data and the
compiled catalog and later literature since 2014, and cross-identified
available galaxy redshift data to obtain RMs and redshifts for 627
pairs. We use these data to study the redshift evolution of RM
differences.  We introduce the rotation measure data in
Section~\ref{sect2} and study the distributions of RM differences of
pairs in Section~\ref{sect3}. { Finally, we discuss our results and
  present conclusions in Section~\ref{sect4} and Section~\ref{sect5},
  respectively.}

Throughout this paper, a standard $\Lambda$CDM cosmology is used, taking 
$H_0=100h$~km~s$^{-1}$Mpc$^{-1}$, where $h=0.7$, $\Omega_m=0.3$ and 
$\Omega_{\Lambda}=0.7$.

\section{Rotation measure data of pairs}
\label{sect2}

We obtain the RM data for a sample of pairs from the NVSS RM catalog
\citep{tss09} and literature \citep[][and afterwards]{xh14}. We search
for real pairs for the two RM datasets separately, since observation
frequencies and resolutions for RM measurements are very
different. The NVSS radio images are visual inspected to ensure
physical pairs.

\subsection{The NVSS RM pairs}

In the NVSS RM catalog, RM data and flux density measurements are
available for 37,543 ``sources''. Here a ``source'' is an independent
radio emission component, while a galaxy can produce a few radio
components, e.g. two unresolved lobes in addition to a compact core of
a radio galaxy. We cross-matched the catalog against itself, and found
1513 source pairs with a flux density ratio $S_{\rm large}/S_{\rm
  small}$ less than 1.5 and an angular separation between $10'$ and
$45''$ (i.e. the angular resolution of the NVSS survey). Flux
densities of real pairs from two lobes of radio galaxies are most
likely to be consistent with each other because of a similar radio
power ejected from the same central black hole. The ratio limit is
therefore used to largely excludes false pairs from two physically
unrelated sources. The maximum separation of $10'$ is set for two
reasons. The first is that it would be difficult to identify
physically related double sources at a larger separation without a
clear connection such as diffuse emission between two sources.
Second, the number of physical pairs at larger separations is small.
In the sample of \citet{vgra19}, only a few pairs have angular sizes
greater than $10'$. The minimum separation was set as being the beam
size $45''$ of the NVSS survey, so that two very close sources can be
just resolved. \citet{vgra19} adopted two times the beam size,
i.e. $1'.5$, while we found the number of physical pairs with
separation $\Delta r < 1'.5$ is more than the twice for pairs with
$\Delta r > 1'.5$, which is important to get pairs for high redshift
galaxies.

\begin{figure}
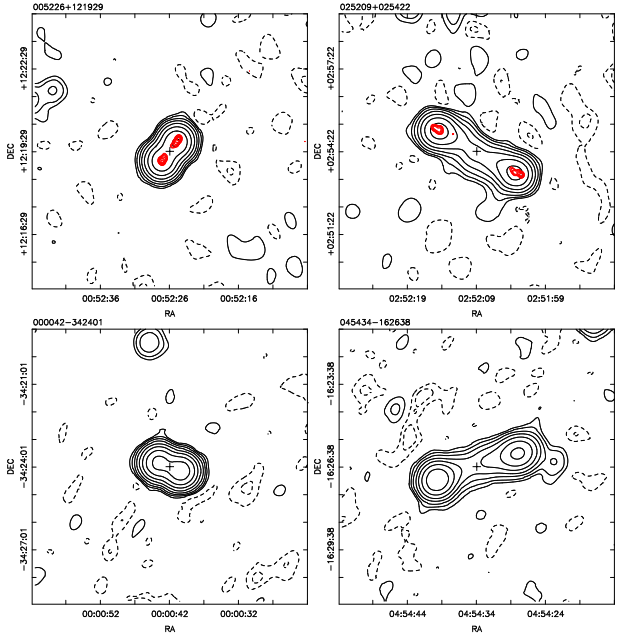

\centering
\includegraphics[angle=-90,width=40mm]{005226+121929.ps} 
\includegraphics[angle=-90,width=40mm]{025209+025422.ps}\\

\includegraphics[angle=-90,width=40mm]{000042-342401.ps} 
\includegraphics[angle=-90,width=40mm]{045434-162638.ps}

\caption{Example images of paired sources from radio galaxies with
  available RMs in the NVSS RM catalog. The top left is the pair of
  J005226+121929 ($\Delta r \simeq 0'.89$); the top right is the pair
  of J025209+025422 ($\Delta r \simeq 3'.05$); the bottom left is the
  pair of J000042-342401 ($\Delta r \simeq 0'.88$); and the bottom
  right is the pair of J045434-162638 ($\Delta r \simeq 3'.05$). The
  pair names here are corresponding to the mean RA and Dec of the
  pair. The top two pairs are located in the FIRST survey area and
  therefore the FIRST contours are shown in red. All contours are
  plotted at levels at $\pm$1, 2, 4, ... mJy beam$^{-1}$, with the
  plus ``+'' indicating the central coordinate of double radio
  sources.}
\label{dbsch}
\end{figure}

Visual inspection was carried out to identify real physical pairs.  We
obtain the NVSS image centered on the mean RA and Dec of each pair,
and make a contour map, as shown in Figure~\ref{dbsch}. For candidates
with angular separations $\Delta r >3'$, the clear presence of fainter
emission connecting the two ``sources'' is the signature for a real
pair, so we get 34 real pairs with $\Delta r >3'$. For pairs with a
smaller angular separation, we check candidates in the survey coverage
area of the VLA Faint Images of the Radio Sky at Twenty centimeters
\citep[FIRST;][]{bwh95} to verify the true pair. With the experience
of classification of real pairs from the NVSS contour maps in the
FIRST area, we extrapolate the method to the sources outside the
survey area of the FIRST. We noticed that physically unrelated pairs
are very scarce at much smaller angular separations
\citep{vgra19,obv+20}. We get 1007 real pairs from the NVSS sources in
total. Four examples of identified real pairs are shown in
Figure~\ref{dbsch}.

For these 1007 pairs, we search for their redshifts of the host
galaxies from several large optical redshift surveys and online
database. First, we cross-match the mean coordinates of RM pairs with
the released spectroscopic redshift of 2.8 million galaxies from Data
Release 16 of the Sloan Digital Sky Survey \citep[SDSS
  DR16,][]{aaa+20}, and we obtain spectroscopic redshift data for
galaxies within 10 arcsec of the given position for 100 pairs. Second,
we get another spectroscopic redshifts from the cross-identification
of galaxies in the 6dF Galaxy Survey Redshift Catalogue Data Release 3
\citep{jrs+09} for 10 pairs. We get photometric redshifts for 227
pairs from the cross-match with the SDSS DR8. For the left sources, we
cross-identified with the NASA/IPAC Extragalactic Database (NED), and
we get redshifts for 64 pairs. In total, we get redshifts and RMs for
401 pairs, as listed in Table~\ref{samplenvss}. The reliability of
such cross-match is about 80\%, as discussed in the
Appendix~\ref{appen}. This is the largest sample of RMs for pairs with
redshifts currently available for the NVSS RM data.

\begin{figure*}
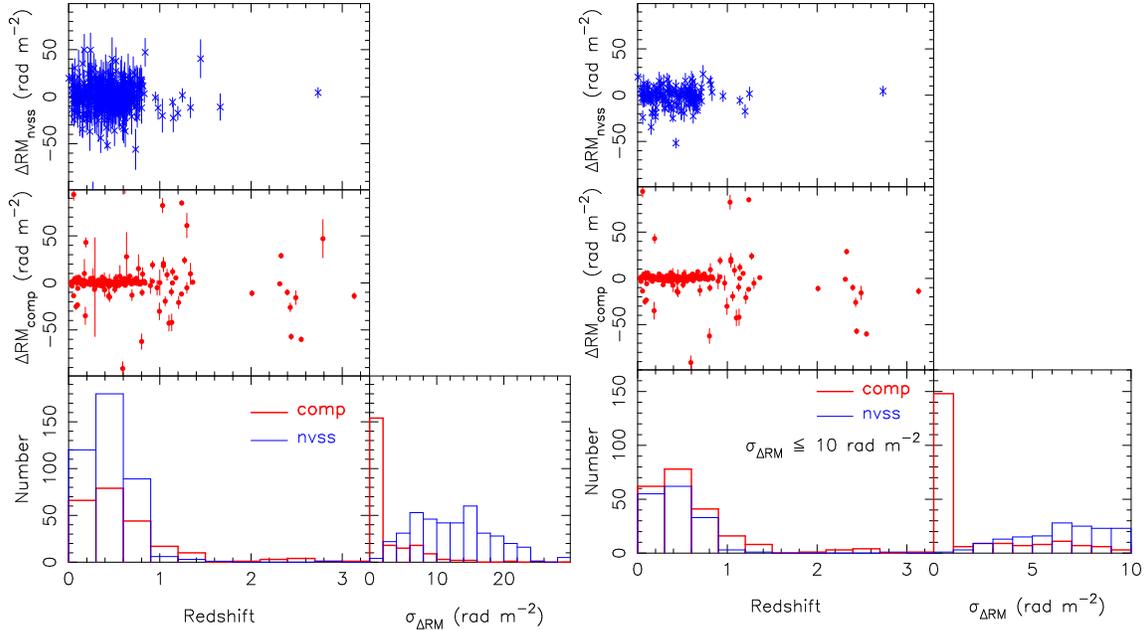

\centering
\includegraphics[angle=-90,width=75mm]{dRMz.ps} 
\includegraphics[angle=-90,width=75mm]{dRMz10.ps} 
\caption{In the left panel, the RM differences $\Delta$RM for 401
  pairs from the NVSS data ({\it top-subpanel}) and for 226 pairs from
  the compiled data ({\it middle-subpanel}) and their histograms ({\it
    bottom subpanel}) against redshift are shown together with the
  histograms for uncertainties $\sigma_{\Delta \rm RM}$.  There are
  2 and 9 pairs with the $\Delta$RM values outside the value range of
  the subpanels for the NVSS and compiled data, respectively.
  The distributions for same data but $\sigma_{\Delta \rm RM}
  \leqslant$ 10 rad~m$^{-2}$ are shown in the right panel.}
\label{dRMz}
\end{figure*}

\subsection{The compiled RM pairs}

In the compiled RM catalog \citep{xh14} and more recently published
literature after then, RMs are available for many pairs, as listed or
presented with radio images in the original references. We inspected
all literature and find 444 double sources as real physical
pairs. Among them 95 pairs have redshifts already listed in the
references or from the NED. For the remaining 349 double sources
without redshifts and known host galaxies, we adopted the same
procedure for redshift search as for the NVSS RM pairs. The central
coordinates of each pair are cross-matched with the SDSS DR16, and we
find spectroscopic redshifts for 40 pairs within 10 arcsec.  No
objects can get the spectroscopic redshift from the 6dF Galaxy Survey
data.  From the catalog of SDSS DR8, we obtain photometric redshifts
for 83 pairs. For the left, we found 8 redshifts from the NED. In
total, we have 226 physical pairs with both RMs and redshifts, as
listed in Table~\ref{samplecomp}. The redshifts for 95 pairs are very
reliable, as marked with '*' in the 10th column, but for the rest 131
pairs, redshift reliability is about 80\%. Notice that the redshifts
of pairs of $z>0.9$ are very reliable, because 34 of the 37 pairs have
redshifts well measured.

\subsection{The RM differences of pairs}

For a physical pair, i.e. the two lobes of a radio galaxy shown as
double radio sources, their radio waves experience almost the same
integration path for the Faraday rotation from their inhabited
environment in front of the radio galaxy to us, as shown in
Figure~\ref{FRIIsch}. The RM difference of a pair indicates mostly the
immediate difference of the magnetoionic medium in their local
environment on a scale comparable to the projected source separation
on the sky plane, i.e. a scale from tens of kpc to a few Mpc, though
we do not know the angle between the line of sight and the pair
connection in 3D. All pairs of sources collected in this work are
unresolved point sources, so that their RMs are produced by almost the
same intervening medium between the source and the observer. The RM
difference $\Delta$RM$=$ RM1 -- RM2 with an uncertainty of
$\sigma_{\Delta \rm RM} = \sqrt{\sigma_{\rm RM1}^2+\sigma_{\rm
    RM2}^2}$ therefore is the cleanest measurements of Faraday
rotation in the ICM, avoiding any additional uncertainties caused by
not-well-measured foreground GRM and by the unknown intergalactic
contributions such as from cosmic webs and galaxy halos. These
unknown uncertainties caused by the foreground of sources are
inherited in all traditional statistics for extragalactic RMs.

The RM difference can be negative or positive as we randomly take one
to subtract the other, so that statistically the zero mean is expected
for a large samples. For our sample the mean of RM difference is
--0.21 and --0.11 rad~m$^{-2}$ for the NVSS and compiled RM pairs,
respectively, which approximate to zero as expected.  The distribution
of $\Delta$RM for two samples of pairs are shown in
Figure~\ref{dRMz}. The RMs, their differences and redshifts of all
these 401 and 226 pairs from the compiled data and the NVSS data are
listed in Table~\ref{samplenvss} and \ref{samplecomp}, respectively,
together with angular separation $\Delta$r, projected linear separation
LS. Only 12 of 401 pairs (3\%) of the NVSS RM sources have redshifts
larger than 0.9, compared with 37 of 226 pairs (16\%) in the compiled
sources. { In the compiled RM data, 34 double sources marked with '--'
in column 11 and 12 have the coordinates for host galaxies but no
coordinates for the two radio lobes, and thus the angular and linear
separations are not available.}

Because the RM uncertainty is a very important factor for the study
of the small RM difference of pairs, and because the formal uncertainty
of the NVSS RM measurements are much larger than those for the compiled
data, the two samples should be analyzed separately. The RM data with
small uncertainties are more valuable to reveal the possible evolution
with redshift, the subsamples with $\sigma_{\Delta \rm RM} \leqslant$
10 rad~m$^{-2}$ are taken seriously here and their distribution is
shown in the right panel of Figure~\ref{dRMz}.

\begin{figure}
\centering
\includegraphics[angle=-90,width=0.8\columnwidth]{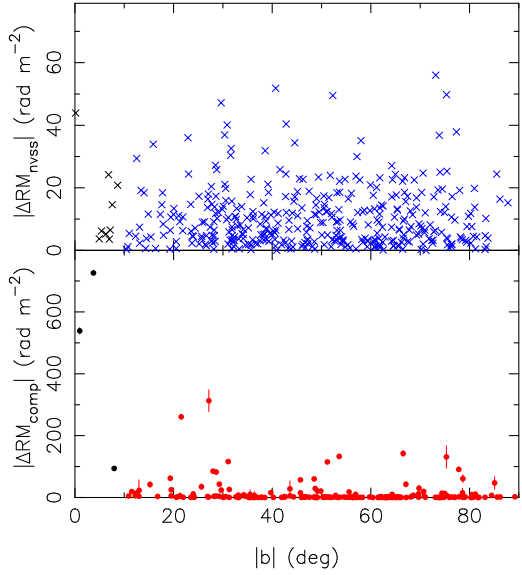} 
\caption{The absolute values of RM difference $|\Delta$RM$|$ of pairs
  from the NVSS data ({\it top panel}) and the compiled data ({\it
    lower panel}) against the Galactic latitudes $|b|$. No apparent
  dependence imply no significant contribution from the ISM. The
  uncertainties of the NVSS RM data are not shown for clarity. }
\label{GB}
\end{figure}

\begin{figure}
\centering
\includegraphics[angle=-90,width=\columnwidth]{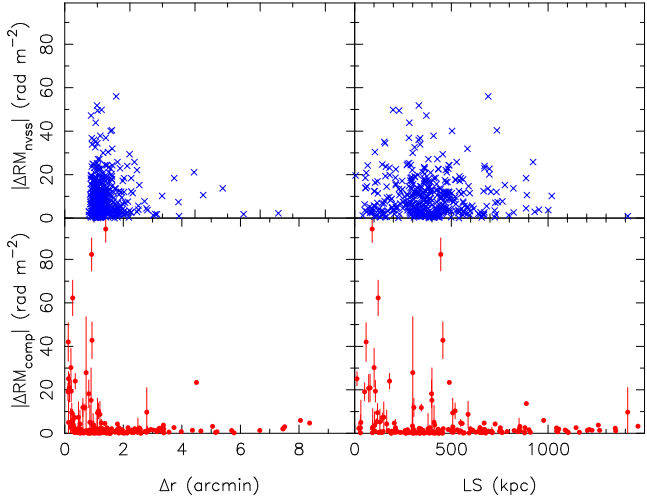} 
\caption{The absolute values of RM difference $|\Delta$RM$|$ of pairs
  for the NVSS data ({\it top panels}) and the compiled data ({\it
    lower panels}) against the angular separation ($\Delta$r) and the
  projected linear separation (LS). The uncertainties of the NVSS data
  are not shown for clarity. A few pairs without the separation values
  or having a RM difference out of the plotted ranges are not shown.}
\label{dRMalsep}
\end{figure}

\section{Large RM difference at high redshifts}
\label{sect3}

Based on this largest samples of pairs with both RMs and redshift data
available so far, we study their evolution with redshift, and check if
the RM difference is related to the separations of two sources.

Figure~\ref{GB} shows the distribution of absolute values of
$|\Delta$RM$|$ of pairs against the Galactic latitude. Because the RM
differences of double sources at low Galactic latitudes may be
contaminated by enhanced turbulence in the interstellar medium when
the radio waves pass through the Galactic plane \citep[e.g.][]{sc86,
  ccsk92,ms96}, we discard 9 NVSS pairs and 3 pairs from the compiled
data at low Galactic latitudes of $|b|<10\degr$, though these few
pairs may not affect our statistics (see Figure~\ref{GB}). A Spearman
rank test demonstrates the absolute $|\Delta$RM$|$ of the NVSS data is
uncorrelated with Galactic latitude, with a correlation coefficient of
$\sim$ --0.004 ($p$-value $\sim$ 0.93). For the pairs from the
compiled data, only a very weak correlation was found from data,
with a correlation coefficient of --0.22 ($p$-value $\sim$ 0.002).
We therefore conclude that the ``leakage'' to the RM differences
from the Galactic interstellar medium can be ignored. 

Figure~\ref{dRMalsep} shows the absolute RM difference as a function
of the angular separation and projected linear separation of two lobes
on the sky plane. For the purpose to explore the magnetic fields in
the intracluster medium, we discard 4 pairs with a LS $\geqslant$
1~Mpc from the NVSS data and 25 pairs from the compiled data, because
these pairs probably impact much less ICM and their differences may
stand more for the RM contribution from the intergalactic medium, given
the typical size of galaxy clusters being about 1~Mpc.
In addition, one pair from a very distant radio galaxy in the compiled
RM data and one pair from the NVSS data have a host galaxy with a
redshift of $z>3$. They are also discarded for the following
statistics.

{ All these discarded pairs are marked with '$\dag$' in the column 13 of
Table~\ref{samplecomp} and \ref{samplenvss}.}  We finally have a very
cleaned 387 NVSS pairs and 197 compiled pairs with a separation of LS
$<$ 1~Mpc, $|b|>10\degr$ and $z<3$ for further analysis.

\begin{figure*}
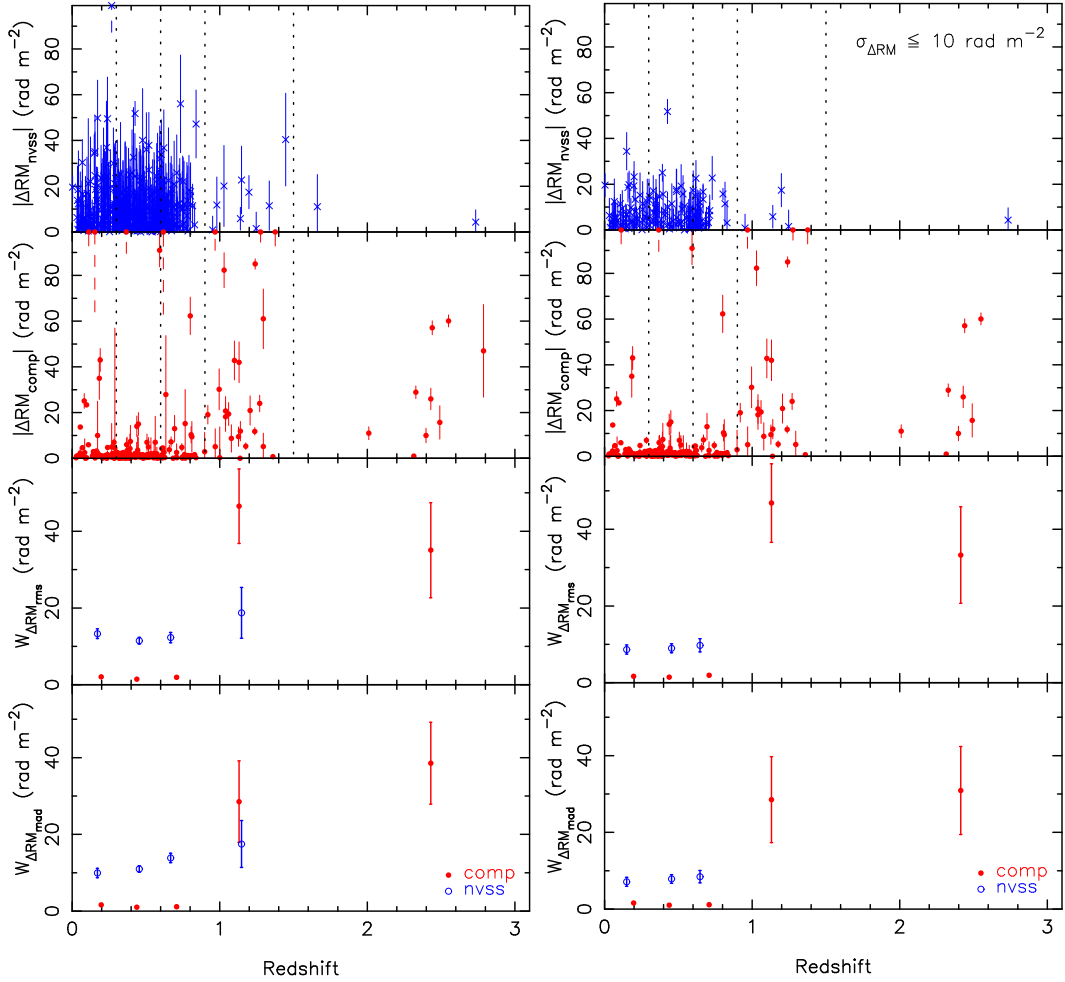

\centering
\includegraphics[angle=-90,width=70mm]{dRMzsta_bl10.ps} 
\includegraphics[angle=-90,width=70mm]{dRMzsta_bl10_rm10.ps} 
\caption{Distribution of absolute values of RM difference $|\Delta \rm
  RM|$ and the data dispersions as a function of redshift for 387 NVSS
  pairs and the 197 pairs of the compiled data with a projected
  separation of LS $<$ 1~Mpc, $|b|>10\degr$ and $z<3$ in the left
  panel. Sources with $|\Delta \rm RM| >$ 100 rad~m$^{-2}$ are plotted
  at top boundary. The vertical dotted lines in the top two rows
  indicate the redshift at $z=$ 0.3, 0.6, 0.9, 1.5. The dispersions of
  the $\Delta$RM distribution are calculated with a Gaussian fitting
  with a characteristic width $W_{\rm \Delta RM}$, or simply taken as
  the median absolute values, as shown in the third and fourth rows of
  panels, respectively. The open circles represent the values from the
  NVSS RM data, and the filled dots stand for values from the compiled
  data, plotted at the median redshift for each redshift range.
  The same plots but for 152 NVSS pairs and 186 compiled pairs with a 
  formal $\Delta$RM uncertainty $\sigma_{\Delta \rm RM} \leqslant$ 10
  rad~m$^{-2}$ are shown in the right.}
\label{dRMzsta}
\end{figure*}

\begin{table*}
\centering
\caption{Statistics of the $\Delta$RM distribution for pairs in redshift bins.\label{dataresult}}
\begin{tabular}{crccccrccc} 
\hline
\multicolumn{1}{c}{ } & \multicolumn{5}{c}{Subsamples from the NVSS RM data}       &     \multicolumn{4}{c}{Subsamples from the compiled RM data}         \\  
Redshift &  No. of   & $z_{\rm median}$   & $W_{\Delta \rm RM_{rms}}$ & $W_{\Delta \rm RM_{mad}}$ &$W_{\Delta \rm RM_{mock}}$~~~&  
            No. of   & $z_{\rm median}$   & $W_{\Delta \rm RM_{rms}}$ & $W_{\Delta \rm RM_{mad}}$~~~~~~\\          
range    &  pairs  &               & (rad~m$^{-2}$) & (rad~m$^{-2}$)  &  (rad~m$^{-2}$) &
            pairs  &   & (rad~m$^{-2}$)  & (rad~m$^{-2}$)   \\
\hline                                                            
\multicolumn{10}{c}{584 pairs with no uncertainty constraint: 387 NVSS RM pairs and 197 compiled RM pairs}\\
\hline
0.0--0.3 &   116     & 0.171         &   13.3$\pm$1.3 & 9.9$\pm$1.2 &10.2$\pm$3.1 &      57     & 0.198         &    2.1$\pm$0.3 & 1.6$\pm$0.3              \\
0.3--0.6 &   174     & 0.455         &   11.5$\pm$0.9 &11.0$\pm$0.8 &10.2$\pm$1.6 &      67     & 0.439         &    1.5$\pm$0.2 & 1.0$\pm$0.2              \\
0.6--0.9 &    86     & 0.668         &   12.3$\pm$1.3 &13.9$\pm$1.2 &10.7$\pm$2.1 &      39     & 0.708         &    2.0$\pm$0.4 & 1.1$\pm$0.4              \\
0.9--1.5 &     9     & 1.148         &   18.7$\pm$6.6 &17.5$\pm$6.1 &  --         &      25     & 1.131         &   46.5$\pm$9.7 &28.5$\pm$10.6             \\
2.0--3.0 &     2     &   --          &         --     &      --     &  --         &       9     & 2.430         &   35.1$\pm$12.4 &38.5$\pm$10.7            \\
0.9--3.0 &    11     & 1.198         &   17.3$\pm$5.5 &17.0$\pm$5.1 &  --         &      34     & 1.222         &   43.7$\pm$7.7  &28.8$\pm$8.2             \\
\hline 
\multicolumn{10}{c}{338 pairs of $\sigma_{\Delta \rm RM} \leqslant$ 10 rad~m$^{-2}$: 152 NVSS RM pairs and 186 compiled RM pairs}\\
\hline
0.0--0.3 &    54     & 0.150         &    8.6$\pm$1.2 & 7.1$\pm$1.2 &7.4$\pm$1.4  &      53     & 0.198         &    1.7$\pm$0.3 & 1.6$\pm$0.2              \\
0.3--0.6 &    60     & 0.454         &    9.0$\pm$1.2 & 7.9$\pm$1.1 &8.2$\pm$0.8  &      66     & 0.438         &    1.5$\pm$0.2 & 1.0$\pm$0.2              \\
0.6--0.9 &    33     & 0.647         &    9.7$\pm$1.7 & 8.5$\pm$1.6 &8.4$\pm$1.9  &      36     & 0.709         &    2.0$\pm$0.4 & 1.1$\pm$0.4              \\
0.9--1.5 &     4     &   --          &         --     &      --     &  --         &      23     & 1.131         &   46.8$\pm$10.2 &28.5$\pm$11.2            \\
2.0--3.0 &     1     &   --          &         --     &      --     &  --         &       8     & 2.414         &   33.3$\pm$12.6 &30.9$\pm$11.5            \\
0.9--3.0 &     5     &   --          &         --     &      --     &  --         &      31     & 1.201         &   43.6$\pm$8.1  &28.5$\pm$8.7             \\
\hline
\multicolumn{10}{l}{$W_{\Delta \rm RM_{mock}}$ denotes the ``intrinsic'' dispersions
  of the NVSS data derived by the mock method in Appendix~\ref{appenB}.}
\end{tabular}
\end{table*}

\subsection{The RM difference versus redshift}

In order to reveal the possible redshift evolution of the small RM
difference caused by the intracluster medium, the $\Delta$RM data have
to be carefully analyzed.

From Figure~\ref{dRMz} and Table~\ref{samplecomp} and
\ref{samplenvss}, we see that the uncertainties $\sigma_{\Delta RM}$
from the NVSS RM measurements have a value between 0 and 25
rad~m$^{-2}$, and those for the compiled RM data are mostly less than
10 rad~m$^{-2}$ and more than half less than 1 rad~m$^{-2}$.
\citet{xh14b} showed that large uncertainties would leak to the
$\Delta$RM distribution. Therefore, we have to study the two samples
of pairs with very different $\Delta$RM uncertainties separately. We
examine two cases, one for the $\Delta$RMs from the whole samples
without a threshold of uncertainty, and the other with the threshold
of $\sigma_{\Delta \rm RM} \leqslant$ 10 rad~m$^{-2}$.

According to number distribution in Figure~\ref{dRMz}, we divide the
samples of pairs in five redshift ranges, z$=$(0.0,0.3), (0.3,0.6),
(0.6,0.9), (0.9,1.5) and (2.0,3.0), and examine the data dispersion in
these ranges as shown in Figure~\ref{dRMzsta}, assuming an
insignificant evolution of RM differences in a given redshift range.
The RM differences of a pair of lobes can be negative or positive, and
for an ideal case of a large sample of the $\Delta$RM values should
follow a Gaussian distribution with the zero mean. The dispersion,
i.e. the width of a Gaussian function $W_{\Delta \rm RM_{rms}}$ and
can be fitted from the real data distribution of $\Delta {\rm RM}$,
through calculating the root mean square (rms) for the $\Delta$RMs:
\begin{equation}
 W_{\Delta \rm RM_{rms}} 
  =\sqrt{\frac{\sum_{i=1,N}(RM1-RM2)_i^2}{N}},
\label{rms}
\end{equation}
here $N$ is the total number of pairs. Alternatively, a more robust
approach is to get the median absolute deviation $\rm W_{\Delta \rm
  RM_{mad}}$, which is good for small data samples and robust in the
presence of outliers \citep[cf.][]{mcs20}.  For our $\Delta$RM data,
the zero mean is expected. Therefore, we consider the median of the
absolute values of the RM difference, i.e.
\begin{equation}
\rm W_{\Delta \rm RM_{mad}}^{\rm ori} = Median(|RM1-RM2|_{i=1,N}).
\label{madfm}
\end{equation}
For a normally distributed data, this can be linked to $W_{\Delta \rm
  RM_{rms}}$ by
%
$  W_{\Delta \rm RM_{mad}} = 1.4826 \times  W_{\Delta \rm RM_{mad}}^{\rm ori}  \simeq W_{\Delta \rm RM_{rms}}$
%
  \citep{llk+13}.

In the redshift ranges with more than five pairs, we calculate the
dispersion of RM differences, $ W_{\Delta \rm RM_{rms}} $ and $
W_{\Delta \rm RM_{mad}} $, see Table~\ref{dataresult} and
Figure~\ref{dRMzsta}.  { Though a large $\Delta$RM is possible for
  embedded double sources contributed from the intracluster medium,
  with a value maybe up to a few hundred rad~m$^{-2}$
  \citep[e.g.][]{ckb01}, a few outliers are cleaned in our statistics
  since they affect the calculation of the dispersion of the main
  stream of data. For the rms calculation, data points scattered away
  from the main distribution by more than three times the standard
  deviation are marked as outliers, and removed iteratively until no
  outliers are marked. The trimmed rms of $\Delta \rm RM$ are
  taken as $ W_{\Delta \rm RM_{rms}} $ for a subsample in a redshift
  bin.} The uncertainty of $ W_{\Delta \rm RM_{rms}} $ is taken as the
standard error for the zero mean, as done by \citet{vgra19}. { For
  the median calculation, the outliers are also cleaned first, and the
  median is found from the remaining $|\Delta {\rm RM}|$, which is
  taken as $ W_{\Delta \rm RM_{mad}}^{ori}$ and then converted to $
  W_{\Delta \rm RM_{mad}}$ with a factor of 1.4826. Its uncertainty is
  taken as being $\sigma_{\left < |\Delta \rm RM_i| \right >}$, the
  error of the estimated mean value of $|\Delta {\rm RM}|$, also with
  a factor of 1.4826. }

The dispersion calculated above in fact includes a ``noise'' term
coming from various uncertainties of RM values. In principle, the
noise term should be discounted from the $\Delta$RM dispersion to get
real astrophysical contributions. For each pair, the noise term can be
expressed from the quadrature sum of the uncertainty of RMs of two
lobes, i.e. for the $i$th pair, the noise $ \sigma_{\Delta \rm RM_i}^2
= (\sigma_{RM1}^2+\sigma_{RM2}^2)_i$. The procedure of noise
subtraction for the dispersion width $\sqrt{ \ W_{\Delta \rm
    RM_{rms}}^2 - \langle \sigma_{\Delta \rm RM_i}^2 \rangle }$
should be carried out under the assumption that the uncertainties in
the observed RMs provide a realistic estimate of the measurement
error. However, RM uncertainties of the NVSS data are underestimated
for most sources \citep{sts11} or probably overestimated for physical
pairs \citep{vgra19}, probably caused by a previously unknown
systematic uncertainty \citep{mgh+10,xh14}. For the compiled RM data,
different estimation methods were used for measurement errors, or
observations with uncorrected ionospheric RM will introduce an extra
RM uncertainty about 3 rad~m$^{-2}$. It is hard to get a realistic
uniformed estimate of the measurement error for the pair sample in
this paper. Fortunately for this work the RM difference $(\Delta \rm
RM)^2$ is concerned, which can largely diminish any systematical
uncertainties which contribute the same amount to the RM measurements
of two closely located sources, though a small unknown amount of noise
leakage still may occur. We found that the $ W_{\Delta \rm RM_{mad}}$
are even much smaller than the average noise power $\langle
\sigma_{\Delta \rm RM}^2 \rangle$, thus no correction of the noise
term is made to dispersion quantities $ W_{\Delta \rm RM_{rms}} $ and
$ W_{\Delta \rm RM_{mad}} $ in Table~\ref{dataresult}.

With these careful considerations above, it is the time to look at
the dispersion of RM differences of pairs as a function of redshift
$z$, with or without a threshold of $\Delta$RM uncertainty for 
the NVSS RM pairs and the compiled RM pairs, respectively.
First of all, the amplitudes of dispersion represented by $ W_{\Delta
  \rm RM_{rms}} $ and $ W_{\Delta \rm RM_{mad}} $ are consistent with
each other within error bars, as shown in Table~\ref{dataresult} and
Figure~\ref{dRMzsta}.
{ Second, for the NVSS RM pairs, no significant variation of the
  dispersion with redshift is seen in both whole sample and the high
  precision sample with $\sigma_{\Delta \rm RM} \leqslant$ 10
  rad~m$^{-2}$, which is consistent with the results for physical
  pairs obtained by \citet{vgra19}.  However, the systematically
  larger dispersion is obtained from the whole sample than that from
  the high precision sample,} which implies the large uncertainty of
the NVSS RM values \citep[a noise term around 10.4 rad~m$^{-2}$ given
  by][]{sch10} significantly affects the dispersion of $\Delta$RM, and
probably buries the small amplitude evolution at low redshifts. This
is a sign of the some noise leakage which cannot be cleaned.
Third, for pairs from the compiled RM data which have very a small
noise, a much larger dispersion appears for pairs of $z>0.9$ in both
samples with/without $\sigma_{\Delta \rm RM}$ threshold setting,
compared to a small dispersion for pairs of $z<0.9$.  { The
  amplitude of dispersion for pairs of $z<0.9$ mostly is less than 2
  rad~m$^{-2}$, but for pairs of $z>0.9$ the dispersion is about 30 to
  40 rad~m$^{-2}$. Even the measurement noise, which is about 5.6/4.7
  rad~m$^{-2}$ at $z>0.9$ without/with $\sigma_{\Delta \rm
    RM} \leqslant 10$~rad~m$^{-2}$ threshold is discounted, the result on larger
  dispersion is not changed. Since the dispersion values for two
  redshift ranges of $z>0.9$ are similar, the data of all pairs in the
  redshift of $0.9<z<3.0$ are therefore jointly analyzed, and the
  uncertainty becomes smaller. The large dispersion for the
  high-redshift pairs of $z>0.9$ is therefore a good detection at
  about a 5-sigma level. }

{ We note that the pairs with a low redshift in the compiled data
  are mainly measured at low frequencies by LOFAR (144~MHz)
  \citep[e.g.][]{obv+20} and MWA (200~MHz) \citep[e.g.][]{rgs+20}.
  Low frequency data may probe the outer part of galaxy clusters or
  poor clusters, hence the dispersion amplitude around 2 rad~m$^{-2}$
  calculated from pairs of $z<0.9$ should read as a lower limit of
  Faraday rotation from the intracluster medium. The dispersion about
  7$\sim$9 rad~m$^{-2}$ estimated from the NVSS RM data with
  $\sigma_{\Delta \rm RM} \leqslant$ 10 rad~m$^{-2}$ should taken as a
  upper limit. The ``intrinsic'' dispersions of the NVSS RM data in
  three low redshift bins at $z<0.9$ are verified by the mock method
  introduced by \citet{xh14b}, see Appendix~\ref{appenB}.

  Based on above results, we conclude that the dispersion of RM
  differences for pairs of $z<0.9$ should be a value in the range of
  2$\sim$8 rad~m$^{-2}$, much smaller than the value of 30$\sim$40
  rad~m$^{-2}$ for high-redshift pairs of $z>0.9$}.

\begin{figure}
\centering
\includegraphics[angle=-90,width=0.7\columnwidth]{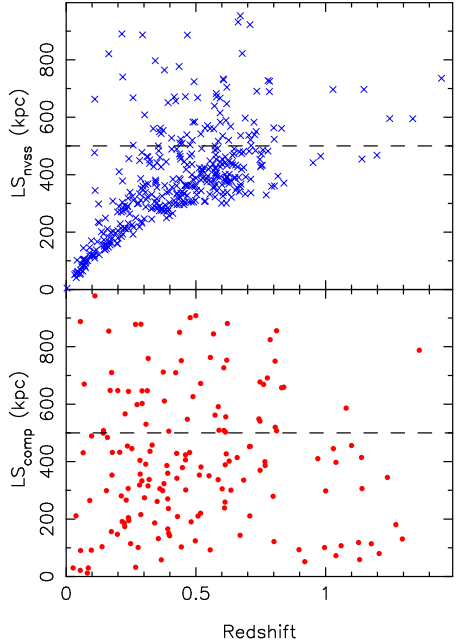} 
\caption{Projected separation of pairs at various redshifts from the
  NVSS sample (\textit{top}) and the compiled data
  (\textit{bottom}). Note that 34 pairs (14 pairs at $z>0.9$) in the
  compiled data are not included since their angular and hence linear
  separations are not available. }
\label{LSz}
\end{figure}

\begin{table*}
\centering
\caption{Statistics of the $\Delta$RM distribution for pairs with a
  separation larger or smaller than 500~kpc.\label{dataresultls500}}
\begin{tabular}{crcccrccc} 
\hline
\multicolumn{1}{c}{ } & \multicolumn{4}{c}{Subsamples from the NVSS RM data}       &     \multicolumn{4}{c}{Subsamples from the compiled RM data}         \\  
Redshift &  No. of   & $z_{\rm median}$   & $W_{\Delta \rm RM_{rms}}$ & $W_{\Delta \rm RM_{mad}}$~~~&  
            No. of   & $z_{\rm median}$   & $W_{\Delta \rm RM_{rms}}$ & $W_{\Delta \rm RM_{mad}}$~~~~~~\\          
range    &  pairs  &               & (rad~m$^{-2}$) & (rad~m$^{-2}$)  &
            pairs  &   & (rad~m$^{-2}$)  & (rad~m$^{-2}$)   \\
\hline                                                            
\multicolumn{9}{c}{pairs with a separation larger than 500 kpc: 76 NVSS pairs and 54 compiled pairs}\\%
\hline
0.0--0.3 &     7     & 0.218         &   10.9$\pm$4.5 &15.7$\pm$3.4       &      15     & 0.199         &    2.6$\pm$0.7 & 2.1$\pm$0.7              \\
0.3--0.6 &    37     & 0.467         &   13.7$\pm$2.3 & 8.5$\pm$2.4       &      19     & 0.467         &    1.1$\pm$0.2 & 1.1$\pm$0.2              \\
0.6--0.9 &    27     & 0.704         &    9.7$\pm$1.9 & 7.1$\pm$1.8       &      18     & 0.754         &    4.0$\pm$1.0 & 2.1$\pm$1.0              \\
0.9--1.5 &     5     & 1.247         &   23.2$\pm$11.6&29.8$\pm$9.6       &       2     &   --          &        --      &    --                    \\
\hline 
\multicolumn{9}{c}{pairs with a separation smaller than 500 kpc: 309 NVSS pairs and 109 compiled pairs}\\%
\hline
0.0--0.3 &   109     & 0.167         &   13.4$\pm$1.3 & 9.7$\pm$1.3       &      34     & 0.210         &    1.7$\pm$0.3 & 1.5$\pm$0.3              \\
0.3--0.6 &   137     & 0.453         &   11.3$\pm$1.0 &11.2$\pm$0.9       &      38     & 0.397         &    1.6$\pm$0.3 & 0.9$\pm$0.3              \\
0.6--0.9 &    59     & 0.653         &   13.3$\pm$1.8 &15.3$\pm$1.6       &      19     & 0.684         &    1.4$\pm$0.4 & 0.6$\pm$0.4              \\
0.9--1.5 &     4     &   --          &         --     &      --           &      18     & 1.114         &   28.1$\pm$6.8 &27.7$\pm$6.9              \\
\hline
\end{tabular}
\end{table*}

\begin{figure*}
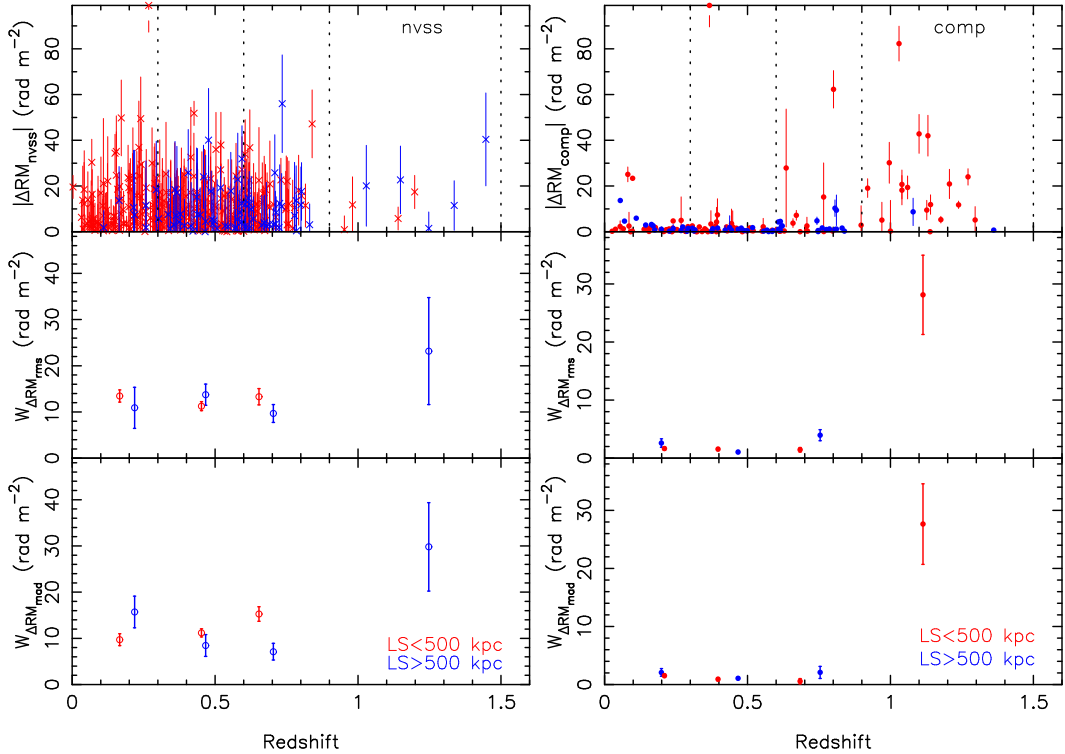

\centering
\includegraphics[angle=-90,width=70mm]{dRMzls5_nv.ps} 
\includegraphics[angle=-90,width=70mm]{dRMzls5.ps} 
\caption{Absolute RM difference ($|\Delta \rm RM|$) distributions and
  their dispersion ($ W_{\Delta \rm RM_{rms}} $ and $ W_{\Delta \rm
    RM_{mad}} $) against redshift for pairs with a separation larger
  and smaller than 500~kpc for the NVSS RM sample ({\it left}) and the
  compiled data sample ({\it right}).  }
\label{dRMzls5}
\end{figure*}

\subsection{The RM difference and projected separation}

Is the significant change of $\Delta$RM dispersion for pairs at
$z>0.9$ biased by the linear sizes of double radio sources or their
separation?  Figure~\ref{LSz} shows the projected separation of pairs
versus the redshift for both the NVSS and compiled RM samples. The
majority of high-redshift pairs ($z>0.9$) in the compiled data have a
separation less than 500~kpc.

As seen in Figure~\ref{dRMalsep}, the absolute values of RM
differences decline to small values when a projected separation is
larger than 1~Mpc, the typical size of a galaxy cluster. The pairs
with such a projected separation greater than 1~Mpc probably lie at a
large angle to the line of sight, and their light paths pass through
much less content of the intracluster medium.

To exam if the larger RM dispersion of high redshift pairs is caused
by different separation, in the following we split the NVSS sample and
compiled data sample into two cases, i.e. the subsamples with a
separation larger or smaller than 500~kpc. In the compiled sample, 34
pairs (14 sources of $z>0.9$) are omitted since the angular and hence
the linear separations are not available though they are probably
smaller than 1~Mpc. Statistics results are shown in
Figure~\ref{dRMzls5} and listed in Table~\ref{dataresultls500}. No
obvious difference of the $\Delta \rm RM$ dispersion can be seen
between two subsamples with a separation smaller and larger than
500~kpc in all three low redshift bins for both the NVSS data and the
compiled data. The dispersion values of subsamples are consistent with
the results derived from the whole sample, which means that the
redshift-dependent dispersion is not caused by different sizes of pair
separation. For high-redshift pairs of $z>0.9$, statistics can be made
for the NVSS subsample with a larger separation than 500~kpc and also
the compiled subsample data with a smaller separation. They both show
a larger dispersion though with different uncertainties. The larger RM
difference is detected at a $4\sigma$ level for the compiled subsample.

\section{Discussion}
\label{sect4}

If the larger RM differences of high-redshift pairs were caused by
intergalactic medium between the pair and us, the larger the
separation between a pair of two lobes, the more likely their radio
waves experience different foreground cosmic filaments and intervening
medium along the lines of sight. That is to say, the larger the
separation of lobe positions, the more likelihood for the larger RM
difference \citep[e.g.][]{omv+19}. 
However, for the compiled RM samples in Figure~\ref{dRMalsep} and
Figure~\ref{dRMzls5}, this is not the case, and the results are just
the opposite, which means that the main RM differences are caused by
the local ICM environment surrounding the double radio sources,
instead of the intervening intergalactic medium in the foreground of a
pair of two lobes. Therefore the RM differences of pairs are excellent
probes for the ICM.

\subsection{Strong magnetic fields in the intracluster medium in the early Universe}

Evidence for larger RM differences for higher-redshift pairs, having
wisely excluded any obvious influence by the Galactic and
intergalactic contributions and also possible dependence on linear
separations of pairs, demonstrates the strong magnetic fields in the
ICM in the early Universe. We can estimate the field strengths in the
ICM from the dispersion of RM differences at the present epoch and at
high redshift.

As mentioned in Section~\ref{sect1}, a pair of lobes are believed to
mainly reside in dense environments of galaxy clusters/groups. Such
dense ambient gas plays a key role in forming Faraday screens which
contributes to the difference between the RM values of the lobes. The
RM asymmetry of a pair of lobes indicates that there probably exists a
large-scale ordered net magnetic fields in the foreground ICM with a
scale of pair separation. Because of turbulent nature for intracluster
magnetic fields, large scale fluctuations ($>$ 100~kpc) should be
responsible for the RM differences of pairs, and a very large outer
scale for turbulent intracluster magnetic fields of $\sim$450~kpc is
possible as being used for modeling of magnetic fields for a giant
radio halo \citep{vmg+10}. The small-scale field fluctuations at a few
kpc could be averaged out over a path length comparable to the
projected separation.

A pair of radio sources in our sample could have any separation and
arbitrary orientations in space. The path difference along the line of
sight of the two lobes may vary from zero to the largest linear size.
Assuming a unidirectional large-scale magnetic field geometry
and a constant electron density in the ambient environs, we get a 
RM difference as being 
\begin{equation}
  \Delta {\rm RM} = 812~n_e B L_{||} \cos{\theta},
\end{equation}
where $L_{||}$ is the separation of the pair (in kpc) projected onto
the line of sight, and $\theta$ is the angle between the magnetic
field direction and the line of sight. For a sample of pairs with the
same separation but random directions of magnetic fields, the mean of
$\Delta$RM is
\begin{equation}
  \left<\Delta {\rm RM}\right> = 812~n_e B L_{||} \int_0^\pi\cos{\theta} \sin{\theta} d\theta \left/\int_0^\pi \sin{\theta} d\theta=0 \right. , 
\end{equation}
and the variance is given by
\begin{equation}
\begin{split}
  \left<(\Delta {\rm RM})^2\right>
&  = (812~n_e B L_{||})^2 \int_0^\pi\cos^2{\theta}\sin{\theta} d\theta \left/\int_0^\pi \sin{\theta} d\theta \right. \\
&  = \frac{1}{3}(812~n_e B L_{||})^2.
\end{split}
\end{equation}
Further more, we consider a pair of sources with a random separation
$L$ along a random orientation $\phi$, i.e. $L_{||} = L \cos{\phi}$,
where $L$ is the size and $\phi$ is the angle between the orientation
and the line of sight. Hence, we expect
\begin{equation}
\begin{split}
  \left<(\Delta {\rm RM})^2\right>
  & = \frac{1}{3}(812~n_e B)^2 \left<L^2\right> \left<\cos^2{\phi}\right> \\
  & =\frac{1}{9}(812~n_e B)^2\left<L^2\right>.
\end{split}
\end{equation}
Here $\left<L^2\right>$ denotes the mean square of the separation of pairs.

The rest-frame RM dispersion of a Faraday screen at redshift $z$ is
expected to be decreased to the observed values by the factor of $(1+z)^2$.
Then we can derive an analytical formulation by assuming the field strength
and the electron density to be constant in the environs around the double
radio sources at redshift $z$, i.e.
\begin{equation}
\begin{split}
  \left<(\Delta {\rm RM})^2\right>
   =\frac{1}{9}812^2\left<L(z)^2\right> \left[\frac{n_e(z)B(z)}{(1+z)^2}\right]^2.
\end{split}
\end{equation}
and finally we get
\begin{equation}
  \begin{split}
    W_{\Delta \rm RM_{rms}}
   & = \left<(\Delta {\rm RM})^2\right>^{1/2}\\
   & = 271~n_{e}(z)B(z)\left<L(z)^2\right>^{1/2} (1+z)^{-2}.
    \end{split}
\label{Wmodel}
\end{equation}
From the Equation~(\ref{Wmodel}), we can derive the magnetic fields in
the ICM if the dispersion of RM difference, electron density $n_{e}$
and the variance of the pair separations $\left<L^2\right>^{1/2}$ at
redshift $z$ are known.

Based on the results shown in Figure~\ref{dRMzsta}, the dispersion of
the RM difference of pairs remains nearly flat at $z<0.9$, with an
amplitude about 2 to 8 rad~m$^{-2}$. We take a typical value being 3.5
rad~m$^{-2}$ to to represent the dispersion at the present time.  For
pairs of $z>0.9$, the dispersion increases to 30 to 40 rad~m$^{-2}$ at
a median redshift of $z=1.1$. We take a typical value being 35
rad~m$^{-2}$ at $z=1.1$. For the variance of the pair separations
$\left<L^2\right>^{1/2}$ at redshift $z$, we take the same typical
value of 350~kpc for pairs at low and high redshifts\footnote{The
  average projected linear separation is 281~kpc and 234~kpc for
  samples of $z<0.9$ and $z>0.9$ with a separation smaller than
  500~kpc, based on the fact that majority of pairs at $z>0.9$
  have small separations and their dispersions are consistent with
  those from the whole sample. Considering random projection
  effect, we estimate the real pair separations should be larger
  by a factor of $\sqrt{2}$=1.4, i.e. 396~kpc or 329~kpc, respectively. }.
At low redshifts, the mean electron density $n_{e}$ in the ICM is
taken to be $4\times 10^{-4}$~cm$^{-3}$, which is obtained by
integrating the $\beta$-model profile of electron density over a
sphere with a radius of 1~Mpc for 12 galaxy clusters
\citep{gdm+10}. According to the Equation~(\ref{Wmodel}), from
$W_{\Delta \rm RM_{rms}} = 3.5$ rad~m$^{-2}$, $n_{e}$ = $4\times
10^{-4}$~cm$^{-3}$ and $\left<L^2\right>^{1/2}$ = 350~kpc at $z=0$, we
can obtain a simple estimation of the magnetic field strength over
this scale as being $B= 0.1 \mu$G at present epoch. At high redshift
$z>0.9$, we do not know the exact properties of the ICM. If we assume
the mean electron density $n_{e}(z)$ at $z>0.9$ is as same as the
density at the present epoch, along with $W_{\Delta \rm RM_{rms}} =
35$ rad~m$^{-2}$, and $\left<L^2\right>^{1/2}$ = 350~kpc as well at
$z=1.1$, the magnetic field would be $B(z) = 4~\mu$G. To get this
value, any field reversals smaller than 350~kpc are ignored. If the
fields reverse at a scale of 30~kpc are considered, the field strength
would be boosted by a factor of $\sqrt{350/30}$, reaching the field
strength of 14~$\mu$G.

\subsection{Implication of strong magnetic fields in the ICM}

The field strength estimated above for the ICM from the pairs at
$z<0.9$, if in the form of a uniform large-scale field geometry, is
0.1 $\mu$G, close to the minimum intracluster magnetic field obtained
by \citet{pjds13}. More tangled fields would have a strength of a few
times stronger. The estimated field strength is smaller than that of
some targeted clusters, such as a few $\mu$G on scales of tens of kpc
in merging clusters and a few 10 $\mu$G in cool core clusters
\citep[see e.g.][]{ct02,vda+19}.  There are two possible reasons.  The
first one, the well-measured RM differences at low redshifts are
predominantly by the RM data with very small uncertainties, which were
mainly measured at low frequencies by LOFAR \citep[e.g.][]{obv+20} and
MWA \citep[e.g.][]{rgs+20}. Those observations at such low frequencies
may probe medium in the outer part of galaxy clusters or poor
clusters, so that the estimated field strength is close to the
large-scale intergalactic magnetic fields around galaxy clusters, as
illuminated by simulations \citep{rkcd08}. In contrast, the small
number of RM data with larger uncertainties and more scattered in the
distribution were mostly observed at 1.4 GHz or higher frequencies,
which are more likely to probe the inner part of galaxy clusters.
Secondly, at low redshifts most powerful radio sources reside in
comparatively sparse environment with few exceptions [e.g. Cygnus A
  \citep{dcp87} and other sources of large RM differences in the
  compiled data], as pointed by \citet{pvc+00}, so that the dispersion
of RM difference is small.  This is supported by the NVSS sample with
a similar small dispersion of RM difference, i.e. upper limit of 7--9
rad~m$^{-2}$ derived by this work and 4.6$\pm$1.1 rad~m$^{-2}$ by
\citet{vgra19}.

The value of uniform intracluster magnetic field strength of 4 $\mu$G
(or $\sim$14 $\mu$G for tangled fields) at $z>0.9$ derived from the RM
difference of pairs is intriguing , as it is comparable to the field
strength of galaxy clusters at low redshifts \citep[see a review
  by][]{han17}, for example a central field strength of 4.7 $\mu$G in
the Coma cluster \citep{bfm+10} and a few microGauss in a sample of
X-ray selected clusters \citep{ckb01,bck16}. This is evidence for
strong organized magnetic fields in galaxy clusters in the early
Universe. If this scenario is correct, it poses a considerable
challenge to theories on the origin of intracluster magnetic fields,
because time available at $z>0.9$ is not sufficient to generate and
align strong magnetic fields on such a large scales. The building-up
of large-scale coherent magnetic fields via the inverse cascade of the
$\alpha-\Omega$ dynamo fields that often works in normal spiral
galaxies cannot operate in galaxy clusters because they do not have an
observed organized rotation. Even if they have, only one or two
rotations at this age of the Universe under slow cluster rotation ($v
\leq 100$ km s$^{-1}$) is insufficient for generation of such a strong
mean field \citep{ct02}.

The origin and the growth of magnetic fields in galaxy clusters are an
enigma. The widely accepted hypothesis is that they are amplified from
much weaker seed fields (either primordial or injected by galactic
outflows) through a variety of processes \citep[see
  review][]{dvbz18}. Simulations show evidence of significant magnetic
field amplification with a small-scale dynamo driven by turbulence and
compression during structure formation \citep{vbbb18,dvbb19}.
Assuming the dynamo growth can start soon after the cluster forms, it
often takes a time-span of several Gyr to amplify magnetic fields to a
few $\mu$G \citep[e.g.][]{dvbb19}. Increasing the Reynolds number can
reduce the time scale for magnetic amplification, but the number is
limited by the efficiency of the transfer of kinetic energy into
magnetic energy. Merger induced shocks that sweep through the ICM or
motions induced by sloshing cool cores may play additional roles in
fast amplification of intracluster magnetic field at high redshifts
\citep{dvbz18}, but not up to such a large scale.  The recent
observations of diffuse radio emission in distant galaxy clusters
\citep{dvb+21} have put a strong limit on the time scale of the
magnetic growth by discovering field strengths of $\mu$G at $z\sim$
0.7. The time available for the amplification in their case is about
3.7~Gyr. Our results is strong evidence for strong magnetic field
strengths at such a large-scale at $z>$ 0.9 and even up to $z\sim$ 2,
comparable to those in nearby clusters, which is a more stringent
constraint for magnetic field generation and evolution.

\section{Conclusions}
\label{sect5}

Faraday rotation measure differences between the two lobes of a sample
of radio galaxies, which is completely free from the Faraday rotation
effect contributed from the interstellar medium inside the Milky Way
and the intergalactic medium between radio galaxies and us, is
significantly large at $z>0.9$, indicating the average intracluster
magnetic fields about 4 $\mu$G (or 14 $\mu$G for tangled fields), in
contrast to the weaker intracluster fields at the present epoch about
0.1 $\mu$G (or 0.3 $\mu$G for tangled fields). Such a strong magnetic
fields in the early universe makes a big challenge on the generation
of cosmic magnetic fields.

More RM data for pairs at high redshift are desired to reach firm
conclusion, since current data sets are not enough in number and have
somehow large measurement uncertainties. Polarization observations for
RMs of a larger sample of double radio sources with a better precision
of RMs should be available soon, which are necessary to further
constrain the evolution of magnetic fields in the ICM.

\appendix
\section{Comparison of redshift data of pairs with those in Vernstrom et al. (2019)
  \label{appen} }

\citet{vgra19} obtained 317 physical pairs of polarized sources at
Galactic latitudes $|b| \geqslant 20\degr$, with a polarization
fraction of $\geqslant2\%$ and an angular separation between $1'.5$
and $20'$ from the NVSS RM catalog \citep{tss09}. These pairs include
not only pairs of lobes, but also separated components of some
multi-component radio galaxies or even multiple RMs within one of the
components (e.g., two RM measurements within one AGN jet or
lobe). Among these 317 pairs, 208 of them have been found to have
spectroscopic or photometric redshifts by performing a redshift search
from the catalogs of \citet{hrg12} and \citet{ki08,ki14} as well as
their own compiled ERG catalog of extended radio galaxies.

We get 1007 real pairs, i.e. double-lobed radio galaxies, based on
visual inspection of the NVSS images with limits on the flux density
ratio $S_{\rm max}/S_{\rm min}$ within 1.5 and the angular separation
between $10'$ and $45''$. The FIRST images are also used to help
classify physical pairs for small angular separations at $\Delta r <
3'$. By crossmatching the mean coordinates of RM source positions with
several large optical redshift surveys and online database, we
obtained  spectroscopic or photometric redshifts for 401 pairs.

\setcounter{figure}{0}
\renewcommand{\thefigure}{A\arabic{figure}}
\begin{figure}
\centering
\includegraphics[angle=-90,width=70mm]{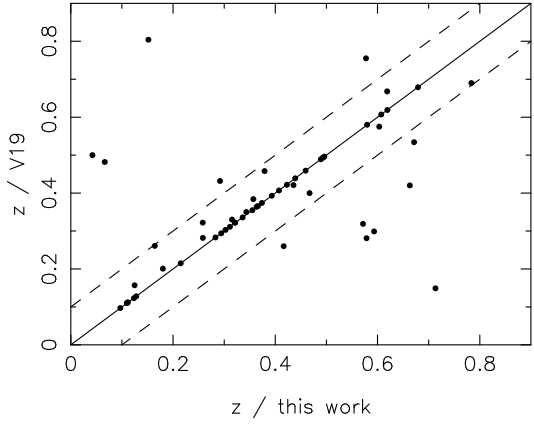} 
\caption{Comparison of redshifts for 56 pairs in
  \citet[][,V19]{vgra19} and our sample. The solid line is the equal
  line, and the dashed lines denote the redshift offset of $\pm$0.1.}
\label{RMzcomp}
\end{figure}

Based on a different approach from \citet{vgra19} to get our RM pairs
and various cross-matches for redshfit, we obtained the large sample
of pairs. Comparing our pairs with those in \citet{vgra19}, we get 56
common samples (see Figure~\ref{RMzcomp}), and 44 of them have
redshifts consistent with each other in $\pm$0.1. The identifications
of host galaxies and redshifts in \citet{vgra19} are probably more
reliable because they considered morphological information, rather
than a simply cross-matching of the central coordinates of pairs with
optical positions as our work. However, our straightforward approach
is a preferable method under the current condition to obtain a large
sample of double radio sources with redshifts. The reliability of our
redshifts, if we take the redshifts of 56 common pairs from
\citet{vgra19} as the standard, should be 44/56, i.e. about 80\%.

\setcounter{figure}{0}
\renewcommand{\thefigure}{B\arabic{figure}}
\begin{figure*}
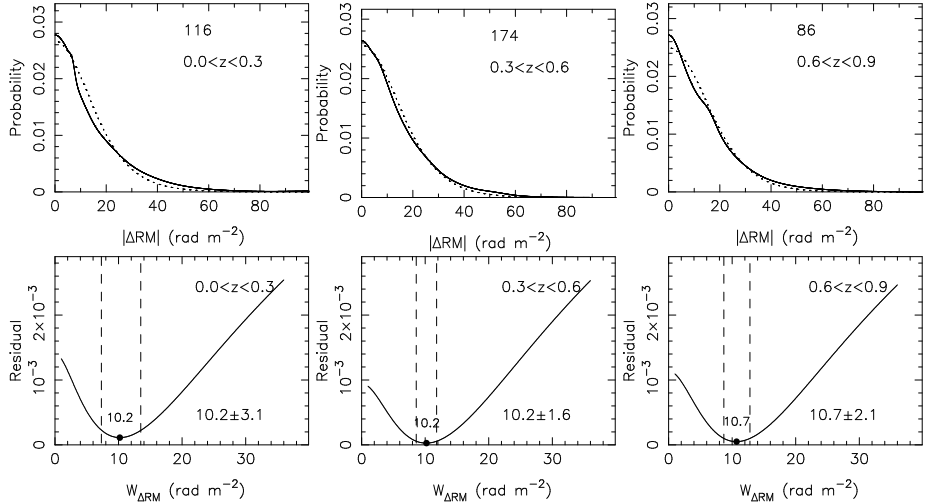

\centering
\includegraphics[angle=-90,width=40mm]{RMz_0.0-0.3.ps}
\includegraphics[angle=-90,width=40mm]{RMz_0.3-0.6.ps}
\includegraphics[angle=-90,width=40mm]{RMz_0.6-0.9.ps}\\
\includegraphics[angle=-90,width=40mm]{RMz_0.0-0.3_res.ps}
\includegraphics[angle=-90,width=40mm]{RMz_0.3-0.6_res.ps}
\includegraphics[angle=-90,width=40mm]{RMz_0.6-0.9_res.ps}
\caption{PDF of $\Delta \rm RM$ values (solid line), compared with that
  of the mock $\Delta \rm RM$ sample with the best distribution dispersion
  $W_{\Delta \rm RM}$ (dotted line), for the whole NVSS sample in three
  redshift bins at $z<0.9$. The fitting residuals against various dispersion
  are plotted in the lower panels, which define the best dispersion and its
  uncertainty at 68 percent probability.}
\label{mocknvss}
\end{figure*}

\begin{figure*}
\centering
\includegraphics[angle=-90,width=40mm]{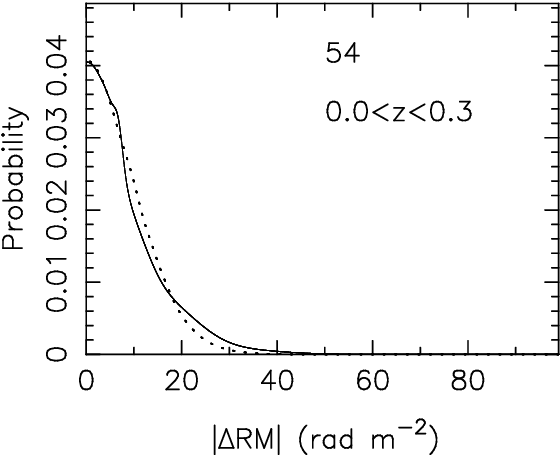}
\includegraphics[angle=-90,width=40mm]{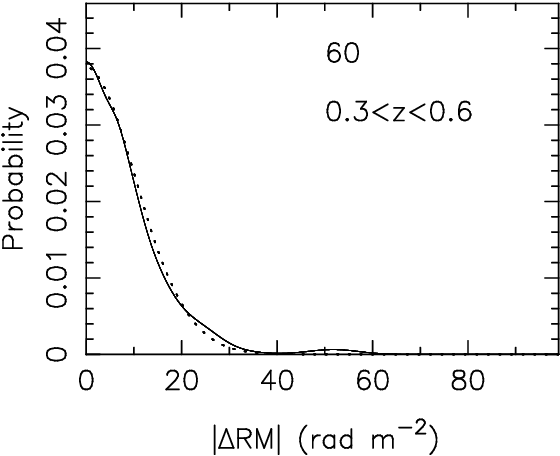}
\includegraphics[angle=-90,width=40mm]{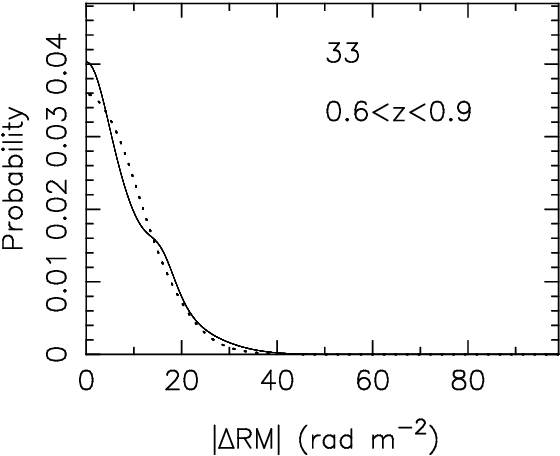}\\
\includegraphics[angle=-90,width=40mm]{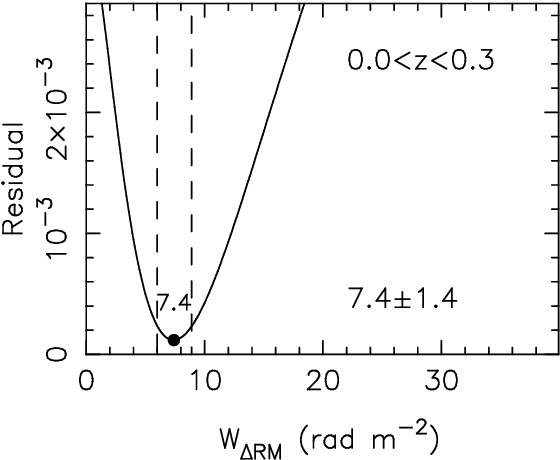}
\includegraphics[angle=-90,width=40mm]{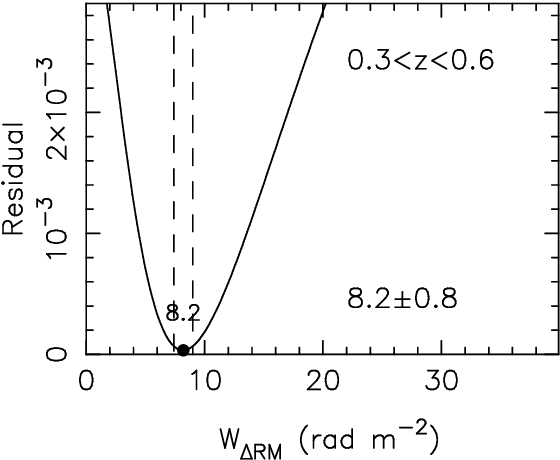}
\includegraphics[angle=-90,width=40mm]{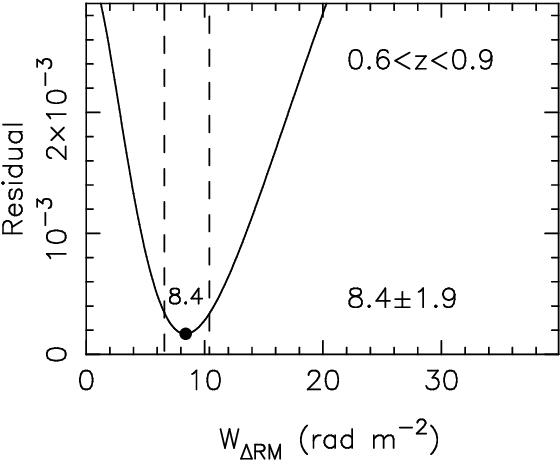}
\caption{Same as Figure~\ref{mocknvss} but for the NVSS sample with
  $\sigma_{\Delta \rm RM} \leqslant$ 10 rad~m$^{-2}$.}
\label{mocknvss10}
\end{figure*}

\section{Application of mock method to the redshift bins at $z<0.9$
  for the NVSS data \label{appenB}}

We tried to use the mock method introduced by \citet{xh14b} to derive
the intrinsic dispersion of a data sample by carefully discarding the
effects from measurement uncertainties. As done in \citet{xh14b},
we use the bootstrap method to obtain the intrinsic dispersion of
RM difference ($\Delta$RM$=$ RM1 -- RM2) distribution given a variety
of uncertainties of $\sigma_{\Delta \rm RM} = \sqrt{\sigma_{\rm RM1}^2+
  \sigma_{\rm RM2}^2}$. It is clear that the probability of a
$\rm \Delta$RM value follows a Gaussian function centred at the
$\rm \Delta$RM value with a width of the uncertainty value, i.e.
\begin{equation}
p(\Delta RM)=\frac{1}{\sqrt{2\pi}\sigma_{\Delta RM_i}}exp \left [ -\frac{(\Delta RM-\Delta RM_i)^2}{2\sigma_{\Delta RM_i}^2} \right ],
\label{normaldis}
\end{equation}
where $\Delta RM_i = (RM1 - RM2)_i$ is the RM difference of the $i$th
of pair in the sample, and $\sigma_{\Delta RM_i}$ is its uncertainty.
Then, we sum the probability distribution function (PDF) for all
$\rm \Delta$RM of a subsample of pairs in a redshift range,
\begin{equation}
P(\Delta RM)=\sum_1^Np(\Delta RM_i).
\label{sump}
\end{equation}
This contains the contributions not only from the intrinsic distribution
width but also the effect of the $\rm \Delta$RM uncertainties. For
an ideal data set without any measurement uncertainty, the PDF follow
a Gaussian distribution with the zero mean and a standard deviation of
$W_{\Delta \rm RM}$, which is the intrinsic dispersion of $\rm \Delta$RM
distribution. Because we make statistics of the absolute difference of
two RMs, we obtain the final PDF only in the positive side by summing
the two half sides of negative and positive profiles. 

Such a method should work well for a larger sample, but does not
work well in the case of a small sample number (e.g. $n<30$) or
mostly a very small uncertainty (e.g. $\sigma<0.1$ rad~m$^{-2}$). For
the NVSS pairs in three redshift bins at $z<0.9$, we got good results.

Following the method introduced by \citet{xh14b}, we generate a mock
sample of $\rm \Delta$RM with a sample size 50 times of the original
$\rm \Delta$RM data with a $\rm \Delta$RM uncertainty randomly taken
from the measured $\rm \Delta$RMs. We then sum the PDF of $\rm
\Delta$RM for the mock data as done for the real data. Finally, by
comparing the two PDFs, $P(\Delta RM)$ and $P_{\rm mock}(\Delta RM)$
as shown in Figure~\ref{mocknvss} and \ref{mocknvss10}, we got the
dispersions at z$=$(0.0,0.3), (0.3,0.6), (0.6,0.9) as being 10.2$\pm$3.1,
10.2$\pm$1.6 and 10.7$\pm$2.1 for the whole sample, and 7.4$\pm$1.4,
8.2$\pm$0.8 and 8.4$\pm$1.9 for the sample with $\sigma_{\Delta \rm
  RM} \leqslant$ 10 rad~m$^{-2}$, also listed in Table~\ref{dataresult}.
The decrease of the derived dispersions of the whole sample to that
of the sample with $\sigma_{\Delta \rm RM} \leqslant$ 10 rad~m$^{-2}$
indicates the leakage of measurement uncertainties to the dispersions.
Therefore, we conclude that the dispersions at $z<0.9$ being 7--9 rad~m$^{-2}$
from the NVSS sample with $\sigma_{\Delta \rm RM} \leqslant$ 10
rad~m$^{-2}$ is the only upper limit of intrinsic dispersion.

\setcounter{table}{0}
\renewcommand{\thetable}{A\arabic{table}}
\startlongtable
\begin{deluxetable*}{rrrcrrrccccccc}
\tablecolumns{14}
\tablewidth{0pt}
\tabletypesize{\footnotesize}
\tablecaption{226 double radio sources with RM and redshift data from the compiled data.\label{samplecomp}} 
\tablehead{
  \multicolumn{4}{c}{Source 1}    &
  \multicolumn{4}{c}{Source 2}    &
  \multicolumn{6}{c}{}     \\
  \hline
  \colhead{RA1}   &  \colhead{Dec1}     & \colhead{RM1}   & \colhead{$\sigma_{\rm RM1}$}  &
  \colhead{RA2}   &  \colhead{Dec2}     & \colhead{RM2}   & \colhead{$\sigma_{\rm RM2}$}  &
  \colhead{Ref.} & \colhead{z} & \colhead{$\Delta$r}  & \colhead{LS}   &
  \colhead{$\Delta$RM} & \colhead{$\sigma_{\Delta \rm RM}$} \\
  \colhead{(deg)}  &  \colhead{(deg)}  & \multicolumn{2}{c}{(rad~m$^{-2}$)} & \colhead{(deg)} &
  \colhead{(deg)}  & \multicolumn{2}{c}{(rad~m$^{-2}$)} & \colhead{}  & \colhead{} & \colhead{(arcmin)}    &
  \colhead{(kpc)}  &\multicolumn{2}{c}{(rad~m$^{-2}$)} 
  }
\startdata
    1.1879 &   12.8131 &   -14.504 &    0.058 &     1.2429 &   12.8000 &   -13.036 &    0.164 & rgs+20  &  0.1430*  &    3.31  &      499 &   -1.468 &   0.174\\
    2.6054 &   33.4822 &   -41.500 &    1.300 &     2.6071 &   33.5031 &   -46.300 &    1.000 & sc86    &  0.7430*  &    1.25  &      549 &    4.800 &   1.640\\
    3.8079 &  -35.4469 &    -2.000 &    7.000 &     3.8079 &  -35.4469 &   -63.000 &   11.000 & opa+17  &  1.2960*  &     --   &       -- &   61.000 &  13.038\\
    4.4926 &  -22.6344 &     8.300 &    2.100 &     4.4926 &  -22.6344 &    19.300 &    1.900 & akm+98  &  2.0100*  &     --   &       -- &  -11.000 &   2.832\\
    4.5386 &   31.0220 &   -76.848 &    0.045 &     4.5350 &   31.0632 &   -77.338 &    0.016 & obv+20  &  0.4996  &    2.48  &      908 &    0.490 &   0.048\\
    7.2502 &   29.7044 &   -62.950 &    0.010 &     7.2732 &   29.7005 &   -60.584 &    0.007 & obv+20  &  0.3882  &    1.22  &      386 &   -2.366 &   0.012\\
    9.6121 &  -38.9961 &   -62.000 &    7.000 &     9.6121 &  -38.9961 &    29.000 &    2.000 & opa+17  &  0.5930*  &     --   &       -- &  -91.000 &   7.280\\
   10.4137 &   41.5087 &   -86.000 &    9.000 &    10.4142 &   41.5109 &   -91.000 &    5.000 & hbb98   &  0.2681  &    0.13  &       32 &    5.000 &  10.296\\
\enddata
\end{deluxetable*}

\startlongtable
\begin{deluxetable*}{rrrcrrrccccccc}
\tablecolumns{14}
\tablewidth{0pt}
\tabletypesize{\footnotesize}
\tablecaption{401 double radio sources with RM and redshift data from the NVSS data. Columns are the same as Table~\ref{samplecomp}.\label{samplenvss}}
\tablehead{
  \multicolumn{4}{c}{Source 1}    &
  \multicolumn{4}{c}{Source 2}    &
  \multicolumn{6}{c}{}     \\
  \hline
  \colhead{RA1}   &  \colhead{Dec1}     & \colhead{RM1}   & \colhead{$\sigma_{\rm RM1}$}  &
  \colhead{RA2}   &  \colhead{Dec2}     & \colhead{RM2}   & \colhead{$\sigma_{\rm RM2}$}  &
  \colhead{Ref.} & \colhead{z} & \colhead{$\Delta$r}  & \colhead{LS}   &
  \colhead{$\Delta$RM} & \colhead{$\sigma_{\Delta \rm RM}$} \\
  \colhead{(deg)}  &  \colhead{(deg)}  & \multicolumn{2}{c}{(rad~m$^{-2}$)} & \colhead{(deg)} &
  \colhead{(deg)}  & \multicolumn{2}{c}{(rad~m$^{-2}$)} & \colhead{}  & \colhead{} & \colhead{(arcmin)}    &
  \colhead{(kpc)}  &\multicolumn{2}{c}{(rad~m$^{-2}$)} 
  }
\startdata
    0.2804 &   23.8897 &   -73.200 &   13.300 &     0.2815 &   23.8724 &   -64.800 &    4.600 &  tss09 &   0.4576 &      1.04 &      363 &    -8.400 &   14.073\\
    0.9972 &  -23.1163 &    15.700 &    6.400 &     1.0258 &  -23.1171 &    19.100 &   12.900 &  tss09 &   0.0970 &      1.58 &      170 &    -3.400 &   14.400\\
    2.6788 &   -6.0596 &    23.800 &   10.300 &     2.6957 &   -6.0677 &     5.300 &    5.800 &  tss09 &   0.0417 &      1.12 &       55 &    18.500 &   11.821\\
    3.3116 &   20.7924 &   -37.900 &   14.900 &     3.3344 &   20.7969 &   -26.100 &    3.300 &  tss09 &   0.2827 &      1.30 &      334 &   -11.800 &   15.261\\
    4.0463 &   32.6553 &    14.600 &   13.100 &     4.0554 &   32.6443 &   -32.600 &    7.000 &  tss09 &   0.8395 &      0.81 &      371 &    47.200 &   14.853\\
    4.5512 &   21.7087 &   -45.900 &   11.900 &     4.5731 &   21.6771 &   -51.600 &    9.700 &  tss09 &   0.3026 &      2.26 &      608 &     5.700 &   15.353\\
    5.0840 &  -20.4925 &   -13.800 &   12.900 &     5.0948 &  -20.4698 &   -11.500 &    4.800 &  tss09 &   0.5450 &      1.49 &      571 &    -2.300 &   13.764\\
    5.1949 &   23.8077 &   -48.900 &   12.400 &     5.1979 &   23.7923 &   -38.100 &   12.700 &  tss09 &   0.5024 &      0.94 &      345 &   -10.800 &   17.750\\
\enddata
\end{deluxetable*}

\begin{acknowledgments}
{ We thank the referee for very careful reading and very detailed
  comments.}
The authors are supported by the National Natural Science Foundation
of China (NNSFC No. 11833009, U2031115, 11988101), and 
the Key Research Program of the Chinese Academy of Sciences
(Grant No. QYZDJ-SSW-SLH021), and also the Open Project Program of the
Key Laboratory of FAST, NAOC, Chinese Academy of Sciences. This research
has made use of the NASA/IPAC Extragalactic Database (NED) which is
operated by the Jet Propulsion Laboratory,California Institute of
Technology, under contract with the National Aeronautics and Space
Administration.
\end{acknowledgments}

\bibliographystyle{aasjournal}
\bibliography{v1a}

\end{document}